\documentclass[twocolumn,english,superscriptaddress]{revtex4}
\usepackage{color}
\usepackage{amsmath}
\usepackage{amssymb}
\usepackage{graphicx}
\usepackage{babel}
\usepackage{subfigure}
\usepackage{epsfig}
\usepackage{float}
\usepackage{mathrsfs}

\begin{document}

\title{Single-atom heat machines enabled by energy quantization}

\author{David Gelbwaser-Klimovsky*}
\affiliation{Department of Chemistry and Chemical Biology, Harvard University,
Cambridge, MA 02138}
\email{dgelbwaser@fas.harvard.edu}

\author{Alexei Bylinskii}
\affiliation{Department of Physics and Department of Chemistry and Chemical Biology, Harvard University,
Cambridge, MA 02138}

\author{Dorian Gangloff}
\affiliation{Cavendish Laboratory, JJ Thomson Ave, Cambridge, UK, CB3 0HE}
\affiliation{Department of Physics and Research Laboratory of Electronics, Massachusetts Institute of Technology,
Cambridge, MA 02139, USA}

\author{Rajibul Islam}
\affiliation{Institute for Quantum Computing and Department of Physics and Astronomy, University of Waterloo, Waterloo, Ontario N2L 3G1, Canada}

\author{Al\'an Aspuru-Guzik}
\affiliation{Department of Chemistry and Chemical Biology, Harvard University,
Cambridge, MA 02138}

\author{Vladan Vuletic}
\affiliation{Department of Physics and Research Laboratory of Electronics, Massachusetts Institute of Technology,
Cambridge, MA 02139, USA}

\begin{abstract}
Quantization of  energy is a quintessential characteristic of quantum systems. Here we analyze its effects on the operation of Otto cycle heat machines  and show that energy quantization alone may alter and increase  machine performance in terms of output work, efficiency, and even operation mode. We show that this difference in performance occurs in machines with inhomogeneous energy level scaling, while quantum machines with homogeneous level scaling behave like classical machines.  Our results demonstrate that quantum thermodynamics   enables the realization of classically inconceivable Otto machines, such as those with an incompressible working substance. We propose to measure these effects experimentally using  a laser-cooled trapped ion as a microscopic heat machine.
\end{abstract}
\maketitle

The discrepancy between classical and quantum mechanics, together with the fast progress on the control of open quantum systems such as ion traps \cite{Rossnagelscience2016,Karpa2013a,Bylinskii2015,Gangloff2015,leibfried2003quantum}, SQUIDS \cite{fornieri2015nanoscale,Bylinskii2015a,Bylinskii2015thesis}, quantum dots \cite{sothmann2014thermoelectric} and molecules \cite{lotze2012driving},
has ignited efforts to clarify the capabilities and thermodynamic limitations of quantum heat machines \cite{gelbwaser2015chapter,kosloff2013quantum,xiao2015construction,roulet2017autonomous}   under quantum effects such as coherences \cite{uzdin2015equivalence,scully2011quantum,gelbwaser2015power}, quantum correlations \cite{correa2013performance,perarnau2015extractable}, quantum statistics of particles \cite{zheng2015quantum}, squeezed thermal baths \cite{rossnagel2014nanoscale,niedenzu2016operation,manzano2016entropy}, many-body effects \cite{jaramillo2016quantum}, and quantized work reservoirs \cite{levy2016quantum,linden2010small,gelbwaser2014heat}. Although these effects may offer classically inaccessible capabilities for machines, there has been no clear evidence that adiabatic quantum machines can outperform their classical counterparts  once all non-equilibrium effects \cite{alicki2015non} and preparation costs are considered \cite{zheng2016cost,torrontegui2017energy}. Among the thermal machines,   one of the most studied  is the  Otto machine \cite{zheng2014work,quan2007quantum}. For this machine, most of the analyses have been limited to potential deformations that homogeneously scale all the energy levels. In this regime, a quantum and a classical heat machine have the same efficiency \cite{quan2007quantum}. The few analyses that consider an inhomogeneous energy scaling \cite{uzdin2014multilevel,quan2005quantum}, have not show a clear advantage of a quantum heat machine over its classical counterpart.

In this Letter, we compare the performance of two \textit{identical} heat machines based on trapped particles (working substance): one governed by classical mechanics and the other by quantum mechanics. We show that the discreteness of energy levels due to quantization, can increase the  efficiency of a heat machine provided that the potential deformation creates an inhomogeneous shift of energy levels. We show that energy quantization can then: \textit{i)}  improve  work extraction, cooling  or  efficiency relative to the classical counterpart,  even reaching the Carnot bound; \textit{ii)}  change the operation
mode, e.g., a heat machine classically expected to operate as a refrigerator, may operate as an engine once energy quantization is considered; \textit{iii)} enable operation at Carnot efficiency even in regimes where classically neither work extraction nor refrigeration are expected.  The origin of the quantum enhanced performance can be traced to the change in relation between temperature and population distribution for adiabatic potential transformations with inhomogeneous energy level shifts.
We emphasize that this analysis relies only on energy quantization and constant level populations in adiabatic potential transformations, and that it does not make use of any hidden resources like non-equilibrium or entangled baths \cite{alicki2015non}.

\begin{figure}
	\centering
		\includegraphics[width=0.5\textwidth]{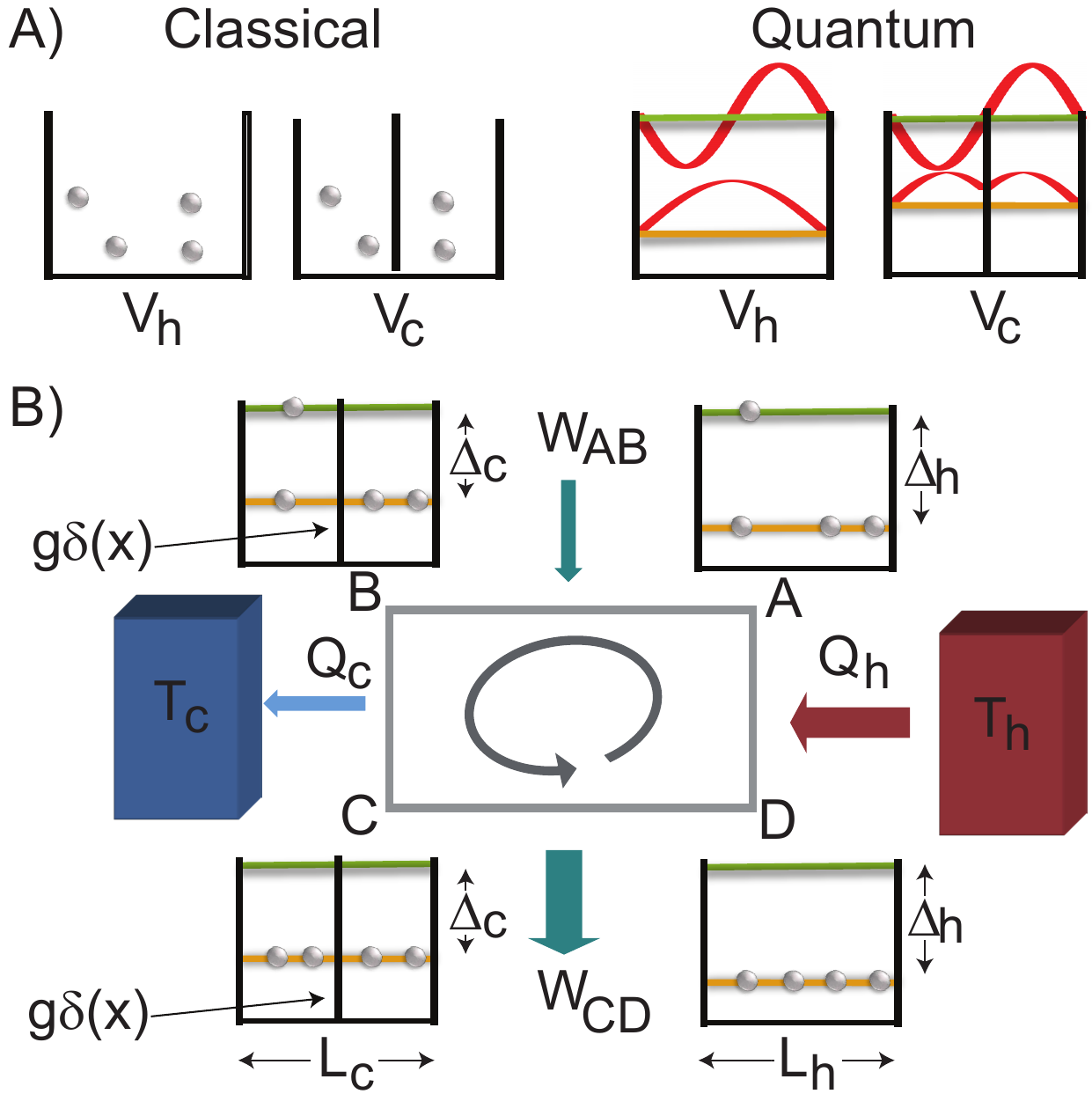}
	\caption{A) The adiabatic introduction of an infinite $\delta-$function  barrier does not change the energy of a  classical ideal gas (left), precluding classical work extraction, but  shifts the energy of the quantum ground state  (right) due to the non-zero amplitude of its wavefunction at the barrier position, enabling quantum work extraction. B) Quantum Otto cycle heat engine using the infinite well potential and the $\delta-$barrier. The cycle is composed of two  adiabatic strokes (connecting states A and B, and C and D) where the potential is adiabatically deformed and the ion does not interact with any thermal bath, and two ``isochoric'' or ``isoparametric'' strokes \cite{zheng2014work} (connecting states B and C, and D and A)  where the potential is kept constant while the ion equilibrates with the cold and the hot bath, respectively see the SI-VII. Work exchange results from the energy shift of the ground state while the excited state remains unshifted (for $L_c=L_h$): work extraction $W_{CD}$ takes place after  thermalization with the cold bath and therefore at high ground-state population, whereas work injection $W_{AB}$ takes place after  thermalization with the hot bath and therefore at lower ground-state population.  The difference between the ground-state populations results in net work extraction $|W_{CD}|>|W_{AB}|$ at constant volume \cite{sandoval2008thermodynamics}.}
	\label{fig:figure1}
\end{figure}

Our results rely on the sensitivity of quantized energies to the boundaries, which classical systems are insensitive to.
 We illustrate this with an example of an Otto engine operated with an ideal gas contained in a  one-dimensional  (1D) infinite  well potential (see Fig.  \ref{fig:figure1}A). The adiabatic introduction of a $\delta-$function barrier at the  center does not alter the volume  nor the classical energy, but by affecting the quantum wavefunctions,  changes the energies of select quantum states. We show that this difference can result in superior performance of quantum heat engines.

Operated as a heat engine,  an Otto machine (see Fig.\ref{fig:figure1}B and SI) transforms incoming heat from the
hot bath, $Q_{h} \geq 0$, into  extracted work, $W<0$, with efficiency $\eta^{en}=\frac{-W}{Q_h}$. It consists of two adiabatic processes where the engine is decoupled from thermal baths, and two isochoric (constant volume) processes where the engine is coupled to two thermal baths at temperatures $T_h$, $T_c$. Note that the efficiency of the Otto engine is not affected by the fact that the isochoric strokes are irreversible (see SI-VII and \cite{leff1975irreversibility,mukhopadhyay2009all}).
 Operated as a refrigerator it consumes work, $W>0$, in order to cool down the
cold bath by extracting heat from it, $Q_{c}>0$, with efficiency $\eta^{ref}=\frac{Q_C}{W}$. We term heater  
the case where the heat flows in its ``natural'' direction from hot to cold, $Q_{h}>0$ and
$Q_{c}<0$, while no work is  extracted, $W\geq0$.

For a classical ideal gas in a uniform potential the compression ratio $r=\frac{\mathcal{V}_c}{\mathcal{V}_h}$ defines the operation mode of the Otto cycle ($\mathcal{V}_{c(h)}$ is the container volume when the ideal gas is at equilibrium with the cold (hot) bath): i)  for $r\leq1$ the machine is a heater; ii) for $1<r<r_{Car} \equiv \left(\frac{T_h}{T_c}\right)^{\frac{1}{\gamma-1}}$  it is an engine; iii) for $r>r_{Car}$ it operates as a refrigerator. 
If run like an engine, the classical efficiency  is 

\begin{equation}
\eta^{en}_{Otto}=1-\frac{1}{r^{\gamma-1}}\leq 1-\frac{1}{r_{Car}^{\gamma-1}}\equiv \eta^{en}_{Car},
\label{eq:ottoef}
\end{equation}
\noindent where  $\gamma=\frac{C_p}{C_v}$ is the specific heat ratio and $\eta^{en}_{Car}$ is the Carnot efficiency limit for an engine. For an incompressible gas, $r=1$ and $\eta^{en}_{Otto}=0$. Classically, a compressible working substance is needed for work extraction, as it has been shown for classical rubber engines \cite{farris1979rubber,mullen1975thermodynamics} and  classical continuum media \cite{paolucci2016continuum,gurtin2010mechanics}. We show below that these paradigms  break down once  energy quantization is included in the analysis.

We first show that if the adiabatic potential deformation during the Otto cycle (from  $V_c$ to $V_h$ and vice versa)  is such that the energy levels scale as $E_{n,h}=qE_{n,c}$, where
$q$ is a positive constant independent of $n$ (homogeneous scaling), then the classical and quantum heat machines always operate in the same mode with the same efficiency. Examples of this type of deformation
are the frequency change of a 1D harmonic trap or the change of length
of a 1D infinite square well potential. Under this assumption, the  work (see SI-I) is 
\begin{gather}
W=\frac{\left(1-q\right)}{q}\int_{qT_{c}}^{T_{h}}C_{v}dT, \label{eq:powercv}
\end{gather}
\noindent where $C_{v}\equiv\frac{\partial\langle H_{h}\rangle_{T}}{\partial T}$
is the heat capacity when the potential is $V_h$, and $\langle\rangle_{T}$ is the expected value
in the thermal Boltzmann distribution at temperature $T$. 
Similarly, the  expressions for the heat transfers are: 
$
Q_h=\int_{qT_{c}}^{T_{h}}C_{v}dT$ and $Q_c=-\frac{1}{q}\int_{qT_{c}}^{T_{h}}C_{v}dT.  
$ 
These expressions  can  also be derived from a completely classical treatment \cite{greiner2012thermodynamics}, where $C_v$ is then the classical heat capacity. Efficiencies, being the ratio between work and heat, do not depend on the heat capacity for  homogeneous energy scaling and under this condition are the same for  classical and for  quantum heat machines. However, the efficiency and even the operation mode can change  when adiabatic potential deformations result in inhomogeneously scaled eigenenergies, $E_{n,h}\neq qE_{n,c}$.

To illustrate this, consider the Otto cycle shown in Fig. \ref{fig:figure1}B where the potential is a 1D infinite  well with variable length $L$, with a thin barrier of width $\epsilon$ that can be adiabatically turned up to a height $V_0$ at the center of the well. Here the energy of a classical particle in thermal equilibrium at temperature $T$ is  $\langle H \rangle\approx\frac{1}{2}k_{B}T+V_0 \frac{\epsilon}{L}e^{-\frac{V_{0}}{k_{B}T}}.$  In the limit of an infinitesimally thin barrier $\epsilon\rightarrow 0$ but constant $g=V_0 \epsilon$, the barrier becomes a $\delta-$function, the energy $\langle H \rangle\rightarrow\frac{1}{2}k_{B}T$ becomes independent of the barrier, and the classical  work output, cooling, and  efficiency correspond to the classical Otto cycle with $r=\frac{L_c}{L_h}$, where $L_{c(h)}$ is the well length at equilibrium with the cold (hot) bath.

\begin{figure} 
	\centering
		\includegraphics[width=0.5\textwidth]{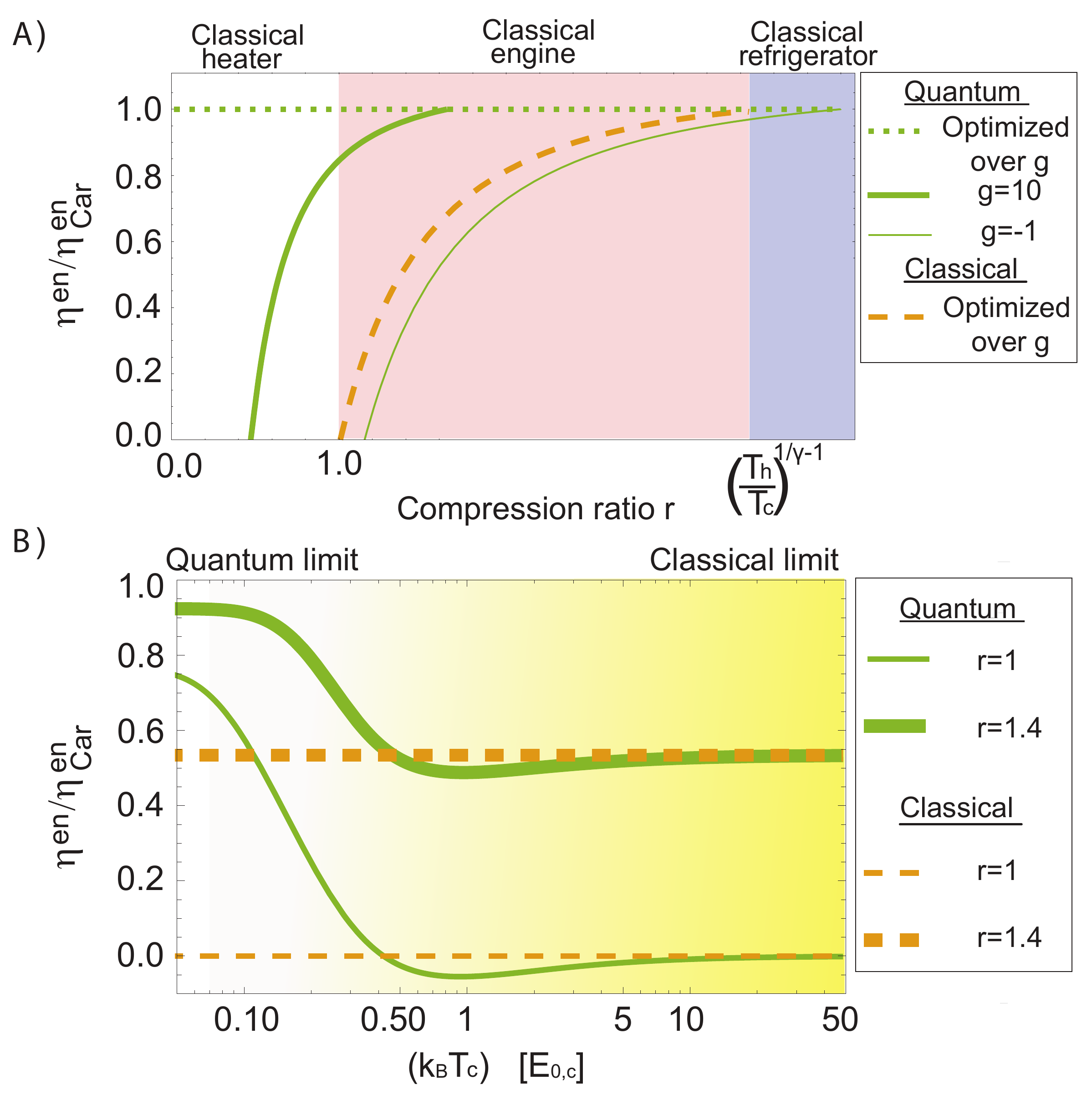}
	\caption{A)   Efficiency normalized to Carnot efficiency for a classical (see SI-III) and quantum (see SI-I) Otto engine with a barrier of the form  $g \delta(x)$ (see Fig. \ref{fig:figure1}B).  The plane is divided into three areas depending on the classical operation mode (see Eq. \eqref{eq:ottoef} and the discussion below it). The temperature ratio is $\frac{T_h}{T_c}=12$, $\gamma=3$,  $g$ is given in units of the critical value $g_{cri}=\frac{2\hbar^2}{m L_c}$, and  units are assumed such that $\frac{2\hbar^2}{m L_h}=1$. B) Classical (dashed) and quantum (continuous) normalized efficiency as function of $T_c$, for  $r=1.4$ (thick) and $r=1$ (thin), and fixed ratio $\frac{T_h}{T_c}$ for an ideal gas. The temperature dependence is a signature of the quantum machine. The number of populated levels depends on the temperature.}
	\label{fig:2dhocomb}
\end{figure}

By contrast, under quantum treatment, the even and odd eigenenergies  are modified differently by introducing the delta barrier, $g \delta(x)$ (see Fig. \ref{fig:figure1}A) \cite{belloni2014infinite}. Odd wavefunctions ($\psi(x)=-\psi(-x)$) remain unperturbed, $E_{n,c}=E_{n,h}$, but $E_{n,c} \neq E_{n,h}$ for   even wavefunctions ($\psi(x)=\psi(-x)$). In this case, the compression ratio  remains $r=1$ and the classical efficiency is zero, while the quantum engine  performs nearly at Carnot efficiency for a high repulsive barrier (see Fig. \ref{fig:2dhocomb}A). Measuring the work extraction during this potential deformation, or other transformation that do not change the ``bulk'' properties of the working substance,  could be used to determine if the working substance is governed by classical or quantum laws. 

Fig. \ref{fig:2dhocomb}B shows that, as one decreases the temperature of the baths at fixed temperature ratio, $T_h/T_c$, the system transitions from a classical regime, where many quantum states are populated, to a quantum regime with  higher efficiency. In the limit of low temperature, where only the two lowest energy levels are appreciably populated, the work extraction condition and efficiency  can be written as (see SI-II)

\begin{equation}
\frac{T_h}{T_c}\geq \frac{\Delta_h}{\Delta_c}>1; \quad\eta=1 -\frac{\Delta_c}{\Delta_h}=1-\frac{1}{r^2} \left( \frac{1}{1-\frac{\Delta E_{c,\delta}}{\Delta_{c}}}\right),
\label{eq:tlscond}
\end{equation}

\noindent where $\Delta_h=E_{1,h}-E_{0,h}$ and $\Delta_c=E_{1,c}-E_{0,c}$ are the energy gaps between excited and ground state when the system is in thermal contact with the hot and cold bath, respectively, and  $\Delta E_{c,\delta}\leq \Delta_c$ is the gap shift produced by the $\delta$ barrier. Eq. \ref{eq:tlscond} shows that the quantum Otto engine may extract work, at Carnot efficiency, for $r=1$ (fixed volume) or any other value of $r$ (see dotted green line in Fig. \ref{fig:2dhocomb}A).  
 Large $g$ even enhances the efficiency at $r<1$, effectively turning a classical heater into an engine. Negative $g$  reduces the efficiency for $r<r_{Car}$, but turns a classically expected refrigerator into a highly efficient quantum engine for $r>r_{Car}$. Carnot efficient quantum engines for any compression ratio can be achieved beyond the two-level approximation. This requires extra control parameters, such as additional $\delta-$barriers, that will ensure that all the energy levels have the appropriate values. In the same way, $g$ could be optimized for reaching maximum work extraction at any compression ratio or for  producing maximum heat extraction, $Q_c$, or refrigeration efficiency $\eta^{ref}$. 

\begin{figure}
	\centering
		\includegraphics[width=0.45\textwidth]{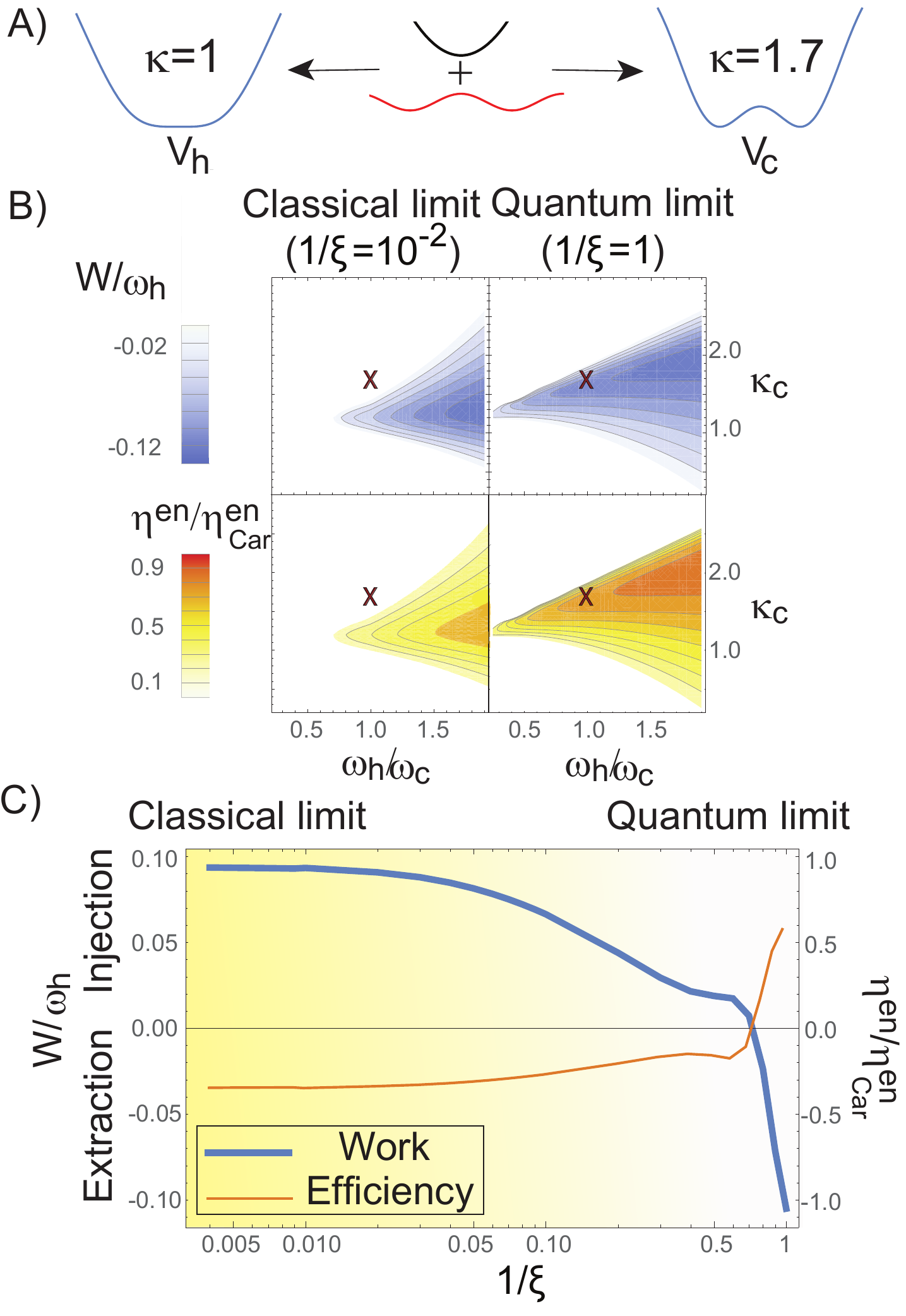}
	\caption{A) The proposed experimental potential (see Eq. \eqref{eq:exppot}).	B) Calculated work  (top) and normalized efficiency (bottom) for the classical (left) and quantum limit (right). Here $\kappa_h=1$ and $\omega_h=1$MHz have been chosen. 		 The white areas  correspond to the heater. The ``X'' indicates the parameters used for   Fig. \ref{fig:powerhfin}C, $\kappa_c=1.7$ and $\omega_c=\omega_h$, where the classical heater operates as a heat engine in the quantum limit.
	C) Calculated work    and   efficiency.  The classical limit is obtained at $\frac{1}{\xi} \rightarrow0$, where the work becomes constant and positive (work injection). In contrast, at the quantum limit, $\frac{1}{\xi}=1$, work is extracted. Cycle parameters: $T_h/T_c = 41.6$, $\bar{n}_c=0.033$.}
	\label{fig:powerhfin}
\end{figure}

The effects of quantization-induced work enhancement and operation mode change can be experimentally tested. To tune a given machine from classical to quantum, one can increase the potentials and temperatures by the same multiplicative factor, $\xi^2$. This effectively decreases the quantization scale relative to the bath temperature and as we show in  SI-IV  this scaling is equivalent to reducing $\hbar$ to zero as $\hbar_{eff} = \hbar/\xi$ (see Fig.  \ref{fig:powerhfin}C).

As a potential experimental platform we consider a trapped, laser-cooled   ion in the combined electrostatic harmonic potential of a Paul ion trap and a  sinusoidal potential of an optical lattice \cite{Karpa2013a,Bylinskii2015,Gangloff2015}. This potential can be used to mimic the infinite well with and without the $\delta-$barrier. The potential has the form 
\begin{equation}\begin{split}
V_i(x)  = m\omega_i^2 a^2 \big( \tfrac{1}{2}(x/a)^2 + 
        \tfrac{\kappa_i}{4\pi^2}\left(1+cos(2\pi x/a)\right) \big), \label{eq:exppot}
\end{split}
\end{equation}

\noindent where $\kappa_i =\omega_{L,i}^2/\omega_i^2$ is the dimensionless parameter controlling the shape of the potential (see Fig. \ref{fig:powerhfin}A),  given by the squared ratio of lattice vibrational frequency $\omega_{L,i}=\sqrt{\frac{2\pi^2 U_i}{m a^2}}$ to the harmonic trap vibrational frequency $\omega_i$. Here $U_i$ is  the depth of the lattice potential. For $\kappa_i=1$,  the potential is a single well while for $\kappa_i>1$, the potential is a double-well, or, equivalently, a single well with a barrier in the middle. The parameter $\kappa_i$ can be tuned  by tuning  $U_i$ (via laser power) and/or the vibrational frequency of the harmonic potential $\omega_i$ (by applying  voltage to the Paul trap electrodes). In Fig. \ref{fig:powerhfin}B we show computational results based on  discrete variable representation (DVR)  calculations \cite{colbert1992novel}  for the work extraction  and efficiency   of the classical  and quantum  versions of the Otto cycle shown in Fig. \ref{fig:figure1}B, but implemented with the  ion-trap potential (Eq. \eqref{eq:exppot}) by adiabatically tuning  $\kappa_i$ and $\omega_i$ in order to generate a double-well and flat-bottom potential. As shown by the marked ``X'', there are parameters for which a classical heater operates as a quantum engine once energy quantization is considered.  Fig. \ref{fig:powerhfin}C shows the  DVR results as function of $\frac{1}{\xi}$ for the parameters of the point ``X'' on Fig. \ref{fig:powerhfin}B. The sign of work flips from positive (work injection) to negative (work extraction) when going  from the classical to the quantum limit. Thus, the turning of a heater into an engine by energy quantization is directly observable 	in a realistic experimental setup.
 
During the isochoric strokes the ion is continuously laser-cooled; at steady-state its temperature is fixed at a stable point where the laser cooling rate balances the heating rate by the environment. The occupation of energy levels then approximately follows a thermal distribution and the system can be considered to be in contact with a thermal bath \cite{An2014a,Meekhof1996}. Contact to a cold thermal bath is achieved by optimizing laser cooling parameters to reduce the steady-state temperature of the ion, whereas contact to a hot thermal bath is achieved by choosing sub-optimal cooling parameters. Raman sideband cooling of $^{174}$Yb$+$ in an $\omega_{L,i}~\sim2\pi\times 10$MHz lattice has been shown to reach near ground-state occupation $\bar{n}\sim 0.1$, and the temperature has been increased controllably by up to a factor of 10 \cite{Karpa2013a,Bylinskii2015,Gangloff2015}. This range could be further increased by reducing external heating sources, and using a narrow optical transition  to precisely measure the motional quantum state populations and ion temperature \cite{Johnson2015}.
The total energy stored in the system $E_T=\sum_n p_n E_n$ at different times can thus be measured via resolved vibrational mode spectroscopy to determine the energy eigenspectrum $E_n$,  and populations  $p_n$ \cite{An2014a}.
 From these measurements, the total work output per cycle can be obtained, and the experiment can be performed in the quantum and classical limits to identify the effects of quantization.

For the adiabatic strokes the laser cooling is disconnected. Perfect adiabaticity has been assumed in the calculation above. In practice, potential deformations during the Otto cycle have to be performed at finite speed, and to avoid excitations that perturb the population distribution, the total adiabatic ramp time must be longer than the inverse of the smallest energy spacing. 
Yet, the ramp time must be  shorter than the thermalization time set by  the background heating in the range $\sim 1-1000\, $ motional quanta per second \cite{eltony2016technologies}. These two conditions can be fulfilled simultaneously for trap vibration frequencies in the MHz range.

We have shown that a quantum Otto engine can be more efficient than its classical  counterpart, but that both are subject to the Carnot limit. This performance difference may be significant since the efficiency of real heat engines \cite{curzon1975efficiency} is limited by the practical difficulty to reach large compression ratios. 
Moreover, we have shown that classically well established paradigms no longer hold in the quantum regime, where  energy quantization allows engines to operate at Carnot efficiency even for compression ratios $r<1$, $r>\left(\frac{T_h}{T_c}\right)^{\frac{1}{\gamma-1}}$ and $r=1$ (fixed volume),  which could enable  the realization of Otto engines with incompressible working substances. These results still hold for a simple model of finite time or imperfect thermalization during the isochoric strokes (see SI-VI), but  more detailed studies are needed to clarify the difference between quantum and classical finite time heat machines.

Since a heat machine operating at given bath temperatures is only characterized by two parameters, the efficiency $\eta$ and the work $W$, it is always possible to construct a classical machine that mimics  a quantum machine with the same $\eta$ and $W$, by choosing a different compression ratio, working fluid, potential deformation, etc. However, here we are interested in differentiating  performance changes based on the quantum/classical nature of the working substance from those originating from other parameter differences. As we have shown, energy quantization, of purely quantum origin, can give rise to a marked difference in performance.

Energy quantization depends on boundary effects, that  generally can be neglected for classical thermodynamic systems, but at the quantum regime allow for work extraction without changing any bulk property of the working substance such as  length for a 1D system (or volume for 3D).

Finally, we have shown that for the studied system non-classical results can be only found when energy levels are inhomogeneously scaled. This regime has rarely been analyzed and requires further investigation. Some potential future research paths include the performance  of other thermodynamic cycles (i.e., Carnot, Stirling, etc), or   the use of a working substance composed of interacting particles or indistinguishable particles (Fermions and Bosons).

\begin{acknowledgments}
 D. G.-K. and A. A.-G. acknowledge the support from the Center for
Excitonics, an Energy Frontier Research Center funded by the
U.S. Department of Energy under award DE-SC0001088 (Energy conversion). V.V. acknowledges support from the NSF (PHY-1505862) and the NSF CUA (PHY-1125846).
\end{acknowledgments}


\newpage
\newpage
\section*{Supplementary information}

\setcounter{equation}{0}
\renewcommand{\theequation}{S\arabic{equation}}

\setcounter{figure}{0}
\renewcommand{\thefigure}{S\arabic{figure}}

\section{Otto cycle}
The  Otto cycle is composed of four strokes that connect different states of the system (A, B, C, D), as follows:
\\
1) At  A, the particle of mass $m$ is in the confining potential $V_{h}$. The Hamiltonian is $H_{h}=\frac{p^2}{2m}+V_{h}(x)$, where $p$ is the momentum operator. The system is at
thermal equilibrium with the hot bath at temperature $T_{h}$, hence the  population of the $n^{\text{th}}$ level is  $P_{n,A}=Z_h^{-1}e^{-\frac{E_{n,h}}{k_BT_h}}$, where $E_{n,h}$ is the $n^{\text{th}}$-level eigenenergy of $H_{h}$ and $ Z_h=\sum_n e^{-\frac{E_{n,h}}{k_BT_h}}$. The potential can be parametrized by a generalized volume $\mathcal{V}_h$ \cite{sandoval2008thermodynamics}.
The system is decoupled from the hot bath and the trap is adiabatically deformed until the potential $V_{c}(x)$ with generalized volume $\mathcal{V}_c$ is obtained at B. The Hamiltonian at B is $H_{c}=\frac{p^2}{2m}+V_{c}(x)$. Adiabaticity ensures that the level populations do not change, $P_{n,B}=P_{n,A}$. The change in energy of the system can be attributed purely to work, $W_{AB}$.
\\
2) Next, the system is coupled to a  cold  thermal bath at temperature $T_{c}$ and it reaches thermal equilibrium at C (see SI-VII). Thus, $P_{n,C}=Z_c^{-1}e^{-\frac{E_{n,c}}{k_BT_c}}$, where $E_{n,c}$ is the $n^{\text{th}}$-level eigenenergy of $H_{c}$ and $ Z_c=\sum_n e^{-\frac{E_{n,c}}{k_BT_c}}$. The trapping potential and its volume do not change; the change in energy of the system can be attributed to heat exchange with the cold bath $Q_c$. 
\\
3) Next, the system is decoupled from the cold bath and the potential is adiabatically
transformed, returning to $V_{h}$ with volume $\mathcal{V}_{h}$ at D.  The level populations do not change,  $P_{n,D}=P_{n,C}$, and all energy exchanged is work, $W_{CD}$.
\\
4) The system is coupled to the  hot bath,  ending at thermal equilibrium with it
at A,  and closing the thermodynamic cycle. The potential is kept constant and the exchanged energy is heat with the hot bath, $Q_h$.

The heat exchanged with the baths is given by the energy difference
between the initial and final states of the isochoric strokes (see SI-VII):
\begin{gather}
Q_{h}=\langle H_{h}\rangle_{A}-\langle H_{h}\rangle_{D}=\sum_{n}E_{n,h}\left(\frac{e^{-\frac{E_{n,h}}{k_{B}T_{h}}}}{Z_{h}}-\frac{e^{-\frac{E_{n,c}}{k_{B}T_{c}}}}{Z_{c}}\right),\nonumber \\
Q_{c}=\langle H_{c}\rangle_{C}-\langle H_{c}\rangle_{B}=\sum_{n}E_{n,c}\left(\frac{e^{-\frac{E_{n,c}}{k_{B}T_{c}}}}{Z_{c}}-\frac{e^{-\frac{E_{n,h}}{k_{B}T_{h}}}}{Z_{h}}\right).\label{eq:heat}
\end{gather}
\indent After completing a cycle, the energy of the system returns to its initial
value. Therefore, by energy conservation, the net work is 
\begin{gather}
W
=-Q_{h}-Q_{c}=
\sum_{n}\left(E_{n,c}-E_{n,h}\right)\left(\frac{e^{-\frac{E_{n,h}}{k_{B}T_{h}}}}{Z_{h}}-\frac{e^{-\frac{E_{n,c}}{k_{B}T_{c}}}}{Z_{c}}\right).\label{eq:work}
\end{gather}
\indent Positive work or heat implies an energy flow into the system and a negative value signifies 
an energy flow out of the system. If
the potential deformation does not change the expected value of the
energies, no work is extracted.

For an homogenous scaling of the energies, $E_{n,h}=qE_{n,c}$, the expression
for the work can be rewritten as 
\begin{gather}
W=\sum_{n}\left(1-q\right)\frac{E_{n,h}}{q}\left(\frac{e^{-\frac{E_{n,h}}{k_{B}T_{h}}}}{Z_{h}}-\frac{e^{-\frac{E_{n,h}}{k_{B}qT_{c}}}}{Z_{c}}\right)=
\frac{\left(1-q\right)}{q} \times\nonumber \\
\left( f_h -f_{q T_c} \right)=
 \frac{\left(1-q\right)}{q} \int_{qT_{c}}^{T_{h}} \frac{d f_T}{dT}dT=\frac{\left(1-q\right)}{q}\int_{qT_{c}}^{T_{h}}C_{v}dT,
\end{gather}
\noindent where $Z_T=\sum_m e^{-\frac{E_{m,h}}{k_{B}T}}$, $f_T\equiv \sum_n E_{n,h} \frac{e^{-\frac{E_{n,h}}{k_{B}T}}}{Z_T}$, $C_{v}\equiv\frac{\partial\langle H_{h}\rangle_{T}}{\partial T}$
is the heat capacity, and $\langle\rangle_{T}$ is the expected value in the thermal Boltzmann distribution at temperature $T$.

\section{Work and efficiency for a two level system}

If only the first two levels are populated Eq. \eqref{eq:work}
can be simplified to

\begin{gather*}
W=(E_{g,c}-E_{g,h}) \times\\\left(\frac{e^{-\frac{E_{g,h}}{k_B T_h}}}{e^{-\frac{E_{g,h}}{k_B T_h}}+e^{-\frac{E_{e,h}}{k_B T_h}}}-\frac{e^{-\frac{E_{g,c}}{k_B T_c}}}{e^{-\frac{E_{g,c}}{k_B T_c}}+e^{-\frac{E_{e,c}}{k_B T_c}}}\right)+\\
(E_{e,c}-E_{e,h})\left(\frac{e^{-\frac{E_{e,h}}{k_B T_h}}}{e^{-\frac{E_{g,h}}{k_B T_h}}+e^{-\frac{E_{e,h}}{k_B T_h}}}-\frac{e^{-\frac{E_{e,c}}{k_B T_c}}}{e^{-\frac{E_{g,c}}{k_B T_c}}+e^{-\frac{E_{e,c}}{k_B T_c}}}\right)\\=
\frac{(\Delta_{h}-\Delta_{c})(e^{-(\frac{E_{g,h}}{k_B T_h}+\frac{E_{e,c}}{k_B T_c})}-e^{-(\frac{E_{g,c}}{k_B T_c}+\frac{E_{e,h}}{k_B T_h})})}{(e^{-\frac{E_{g,h}}{k_B T_h}}+e^{-\frac{E_{e,h}}{k_B T_h}})(e^{-\frac{E_{g,c}}{k_B T_c}}+e^{-\frac{E_{e,c}}{k_B T_c}})},
\end{gather*}

\noindent where $\Delta_{i}=E_{e,i}-E_{g,i}.$ Therefore, the condition
for work extraction, $W<0$, is 

\[
\frac{T_{h}}{T_{c}}>\frac{\Delta_{h}}{\Delta_{c}}>1.
\]

In a similar way, the heat exchanged with the hot bath is 

\[
Q_{h}=\Delta_{h}\frac{e^{-(\frac{E_{g,h}}{k_B T_h}+\frac{E_{e,c}}{k_B T_c})}-e^{-(\frac{E_{e,c}}{k_B T_c}+\frac{E_{e,h}}{k_B T_h})}}{(e^{-\frac{E_{g,h}}{k_B T_h}}+e^{-\frac{E_{e,h}}{k_B T_h}})(e^{-\frac{E_{g,c}}{k_B T_c}}+e^{-\frac{E_{e,c}}{k_B T_c}})},
\]

\noindent and the efficiency is 

\[
\eta^{en}=-\frac{W}{Q_{h}}=1-\frac{\Delta_{c}}{\Delta_{h}}.
\]

For the cycle shown in figure 1B in the main text, $\Delta_h=\frac{3\hbar^2 \pi^2}{2m L_h^2}$ and $\Delta_c= \Delta_{c,box}+\Delta E_{c,\delta}$, where  $\Delta_{c,box}=\frac{3\hbar^2 \pi^2}{2m L_c^2}$. Therefore,

\begin{gather}
\frac{\Delta_c}{\Delta_h}=\frac{\Delta_{c,box}}{\Delta_h} \left(\frac{\Delta_c}{\Delta_{c,box}}\right)= \notag \\
\frac{1}{r^2}\left(\frac{\Delta_c}{\Delta_c-\Delta E_{c,\delta}} \right)= \frac{1}{r^2}\left(\frac{1}{1-\frac{\Delta E_{c,\delta}}{\Delta_c}}\right),
\label{eq:}
\end{gather}
where we have used the fact that $r^2=\frac{\Delta_h}{\Delta_{c,box}}=\frac{L^2_c}{L^2_h}$. From here the right side of Eq. 3 in the main text is derived.

\section{Thermodynamic calculations for the classical heat machine}

There are multiple alternative methods to calculate the $Q_h$, $Q_c$ and $W$ for the classical heat machine studied in the main text. All of them give the same results:

\begin{enumerate}
	\item Doing the quantum calculation using Eqs. \eqref{eq:heat} and \eqref{eq:work}, and effectively reducing $\hbar$ until the result converges. In the studied cases the convergences was obtained for $\hbar_{eff}/\hbar=10^{-2}$;
	\item Considering the same scaling for the potential and temperatures, $V_i\rightarrow \xi^2V_i$ and $T_{c(h)}\rightarrow \xi^2T_{c(h)}$
and taking the limit $\xi \rightarrow \infty$;
	\item In the case of the infinite square well,  the $\delta-$barrier does not change the energy at the classical limit, the standard Otto cycle calculation can be used, neglecting the $\delta-$barrier.

\end{enumerate}

\section{Experimental simulation of the classical limit}
In this section we show that the classical limit of the extracted work,
\eqref{eq:work}, is equivalent to the work obtained after scaling the potential
and the temperature. A similar proof can be used for the heats. 

In order to find the classical limit, in the Schrodinger equation
$\hbar$ is replaced by $\hbar_{eff}=\frac{\hbar}{\xi}$, where $\xi$
is scaling parameter in the range between $1$ and $\infty$.  The Schrodinger equation is 

\begin{gather}
\left[-\left(\frac{\hbar^{2}}{\xi^{2}}\right)\frac{1}{2m}\frac{\partial^{2}}{\partial^{2}x}+V(x)\right]\psi_{n}(x)= \notag\\
E_{n}\left(\hbar_{eff}=\frac{\hbar}{\xi},V(x)\right)\psi_{n}(x),
\end{gather}
where the eigenenergies depend on $\hbar_{eff}$ and on $V(x)$. By
multiplying both side by $\xi^{2},$

\begin{gather}
\left[-\frac{\hbar^{2}}{2m}\frac{\partial^{2}}{\partial^{2}x}+\xi^{2}V(x)\right]\psi_{n}(x)= \notag \\\
\xi^{2}E_{n}\left(\hbar_{eff}=\frac{\hbar}{\xi},V(x)\right)\psi_{n}(x).
\end{gather}

\noindent Therefore, we conclude that,

\begin{gather}
E_{n}\left(\hbar_{eff}=\hbar,\xi^{2}V(x)\right) = \notag \\
 \xi^{2}E_{n}\left(\hbar_{eff}=\frac{\hbar}{\xi},V(x)\right).\label{eq:scalrel}
\end{gather}

The work is a linear combination of terms of the form,

\begin{gather}
\sum_{n}E_{n}\left(\hbar_{eff}=\frac{\hbar}{\xi},V_{a}(x)\right)\frac{e^{-\frac{E_{n}\left(\hbar_{eff}=
\frac{\hbar}{\xi},V_{b}(x)\right)}{k_{B}T}}}{Z_{T,\hbar_{eff}=\frac{\hbar}{\xi},V_{b}(x)}},
\end{gather}

\noindent where $V_{a}(x)$ and $V_{b}(x)$ may be the same or different potentials
and $Z_{T,\hbar_{eff}=\frac{\hbar}{\xi},V_{b}(x)}=\sum_{n}e^{-\frac{E_{n}(\hbar_{eff}=\frac{\hbar}{\xi},V_{b}(x))}{k_{B}T}}$
. Using Eq. \eqref{eq:scalrel} we get 

\[
\sum_{n}\frac{E_{n}\left(\hbar_{eff}=\hbar,\xi^{2}V_{a}(x)\right)}{\xi^{2}}\frac{e^{-\frac{E_{n}\left(\hbar_{eff}=\hbar,\xi^{2}V_{b}(x)\right)}{k_{B}\xi^{2}T}}}{Z_{\xi^{2}T,\hbar_{eff}=\hbar,\xi^{2}V_{b}(x)}}.
\]

\noindent From this we conclude that 
 
\begin{gather}
W(\hbar_{eff}=\frac{\hbar}{\xi},V_{a}(x),V_{b}(x),T_{h},T_{c})= \notag \\
\frac{W(\hbar_{eff}=\hbar,\xi^{2}V_{a}(x),\xi^{2}V_{b}(x),\xi^{2}T_{h},\xi^{2}T_{c})}{\xi^{2}}.
\end{gather}

The classical
limit is obtained for large $\xi$ when $W(\hbar_{eff}=\frac{\hbar}{\xi},V_{a}(x),V_{b}(x),T_{h},T_{c})$ becomes a constant as function of $\hbar_{eff}=\frac{\hbar}{\xi}$. 
Thus, by scaling the potential and the temperature by a large factor,
$\xi\rightarrow\infty,$ it is possible to experimentally simulate
the classical limit, $\hbar\rightarrow0$.

The  scaling of a potential and the temperature has been achieved in ion traps setups \cite{Karpa2013a,Bylinskii2015,Gangloff2015,Bylinskii2015a,Bylinskii2015thesis}.  Therefore, we consider them as the ideal platform to test the  classical and quantum limit of the same heat machine. 

\section{Carnot limit}
In this section we prove that the efficiency of the Otto quantum heat machine is bounded
by the Carnot limit, $\eta^{en}_{car}=1-\frac{T_{c}}{T_{h}}$. We focus
on the heat engine efficiency but the bounds for the performance of
a refrigerator can be derived in the same way. 
The efficiency of a heat engine is

\[
\eta^{en}=\frac{-W}{Q_{h}}.
\]

Work extraction requires $W<0$ and $Q_{h}>0$. The expression
for the   heat and the work are given by Eqs. \ref{eq:heat} and  \ref{eq:work} on the SI-I. As a first step, assume that the work and heat are produced
by a single level,

\begin{gather}
W_{n}=\left(E_{c,n}-E_{h,n}\right)\left(\frac{e^{-\frac{E_{h,n}}{k_{B}T_{h}}}}{Z_{h}}-\frac{e^{-\frac{E_{c,n}}{k_{B}T_{c}}}}{Z_{c}}\right);\notag \\
Q_{h,n}=E_{h,n}\left(\frac{e^{-\frac{E_{h,n}}{k_{B}T_{h}}}}{Z_{h}}-\frac{e^{-\frac{E_{c,n}}{k_{B}T_{c}}}}{Z_{c}}\right).
\end{gather}

Work extraction  requires $\frac{E_{h,n}}{T_{h}}<\frac{E_{c,n}}{T_{c}}$, otherwise, $W>0$. Therefore, the single level efficiency is bounded, 

\begin{equation}
\eta^{en}_{n}=1-\frac{E_{c,n}}{E_{h,n}}\leq1-\frac{T_{c}}{T_{h}}\label{eq:carnot}.
\end{equation}

Next we consider two levels, $n$ and $m$.  We prove that the efficiency in the case of two
levels can not be greater that the efficiency of a single level and
therefore the two level case is also bounded by the Carnot limit.
Assume that the efficiency of the two levels is greater than the single
level efficiency, 

\begin{equation}
\frac{-W_{n}-W_{m}}{Q_{h,n}+Q_{h,m}}>\frac{-W_{n}}{Q_{h,n}}.\label{eq:double}
\end{equation}

Work extraction requires    $Q_{h,n}+Q_{h,m}>0$, thus $Q_{h,n}$ or $Q_{h,m}$ should be positive. Without loss of generality we assume $Q_{h,n}>0$ and $\eta^{en}_{n}\geq\eta^{en}_{m}$. Hence, Eq. \eqref{eq:double} can be rewritten as 

\begin{equation}
\frac{-W_{m}}{Q_{h,m}}>\frac{-W_{n}}{Q_{h,n}}.
\label{eq:cont}
\end{equation}

Equation \eqref{eq:cont} contradicts the assumption $\eta^{en}_{n}\geq\eta^{en}_{m}$. Thus, the inequality on Eq. \eqref{eq:double} does not hold. This can be generalized for a multilevel system. Therefore,  the efficiency of a multilevel system
can not be greater than the highest single-level efficiency. The latter,
and therefore the whole multilevel  efficiency, is bounded by
the Carnot limit, (see Eq. \eqref{eq:carnot}).

\section{Finite time Otto cycle}
We consider a simple model of a finite time Otto cycle where the system
does not fully equilibrate with the thermal baths during the isochoric
strokes. Instead, we assume that the isochoric strokes are interrupted before
equilibration and the system ends in a mixture of the initial state
and the thermal state, i.e.,
\[
\rho_{no-th}= p\rho_{0}+(1-p)\rho_{th},
\]
 where $\rho_{o}$ is the initial state of the system at the beginning
of the isochoric stroke and $\rho_{th}$ is the equilibrium state
it would have reached after  infinite time. $p$ is a constant that
represents the degree of thermalization and goes from 0 for a fully
equilibrium state, to 1 for a state that did not thermalize at all.
For simplicity we assume that the degree of thermalization is symmetric,
i.e., it is the same for the two isochoric strokes. The heat transfer
from the hot bath  is
\begin{gather}
Q_{h}^{inc}=\langle H_{h}\rangle_{\rho_{no-th}}-\langle H_{h}\rangle_{\rho_0}= \nonumber\\
Tr[H_{h}(p\rho_{0}+(1-p)\rho_{th})]-Tr[H_{h}\rho_{0}]=\nonumber\\
Tr[H_{h}(p\rho_{0}+(1-p)\rho_{th})]-Tr[H_{h}(p+1-p)\rho_{0}]=\nonumber\\
(1-p)\left(Tr[H_{h}\rho_{th}]-Tr[H_{h}\rho_{0}]\right)=(1-p)Q_{h},
\end{gather}
 where $Q_{h}$ is the heat exchanged if the system fully thermalizes
(see Eq. \ref{eq:heat}). A similar expression is found for $Q_{\text{c}}^{inc}.$
Therefore, the work extracted during this cycle is 
\[
W^{inc}=-Q_{h}^{inc}-Q_{c}^{inc}=-(1-p)W,
\]
but the efficiency, being the ration between $W^{inc}$ and $Q_{h{c}}^{inc},$
remains the same, and the operation mode of the heat machine does
not change. Therefore, if the degree of thermalization is symmetric,
the effects shown in the main text do not change. More complex finite
time cycles could be considered, but they are out of scope of this
paper and are left for  future works.

\section{Irreversibility of the isochoric process}

During the isochoric process the working substance is coupled to a thermal bath which not necessary is close to the working substance temperature which makes the process irreversible.  In addition, in the quantum system, the state after the adiabatic step prior to the isochoric process is in general not a thermal state.
However, the fact that the isochoric process is irreversible does not change the efficiency of the Otto engine. The efficiency is defined in terms of the work done by the engine, and the heat received by the engine from the hot reservoir \cite{greiner2012thermodynamics}. No work is done during the isochoric process, and therefore the work done is not affected by irreversibility. Furthermore, the heat exchanged during each isochoric process is given by the internal energy difference of the working substance between the beginning and the end of the isochoric process, and is path independent. Therefore the heat exchanged by the engine and the reservoir is also not affected by irreversibility. Consequently the efficiency of the Otto engine is not affected by the irreversibility of the isochoric process. This is also the conclusion reached after Eq. 17 of Ref. \cite{mukhopadhyay2009all}. Note that this argument does not hold for other processes where work is exchanged while the system is coupled to a bath.

\end{document}